# What If Memory Information is Stored Inside the Neuron, Instead of in the Synapse?

James Tee, *Member, IEEE*, and Desmond P. Taylor, *Life Fellow, IEEE*

*Abstract*—Memory information in the brain is commonly believed to be stored in the synapse. However, a recent groundbreaking electrophysiology research has raised the possibility that memory information may actually be stored inside the neuron itself. Drawing on information theory and communications system engineering perspectives, we examine the problem of how memory information might be transmitted reliably between neurons. We identify 2 types of errors that affect neuronal communications (i.e., channel error and source error), along with plausible error mitigation solutions. We confirm the feasibility of these solutions using simulations. Four alternative hypotheses of the synapse's function are also proposed. We conclude by highlighting some research directions, along with potential areas of application.

*Index Terms*—brain, memory, information, synapse, coding, communications, computations.

## I. Introduction

Conventional wisdom has it that memory information in the brain is stored in the synapse [1][2]. This is a foundational basis for computational neuroscience [3][4]. The synapse is also the underlying basis for artificial intelligence (AI) approaches that we know of today. For example, deep neural networks based on synaptic weights [5], which have been successful in mastering the game Go [6], even without any human guidance at all [7]. More recently, these neural networks were successful in solving protein folding problems [8]. These impressive feats seemingly reinforce the Hebbian synaptic hypothesis on how the brain learns [9].

Despite its popularity, the Hebbian (plastic) synaptic hypothesis still lacks conclusive proof that the brain's memory information resides in the synapse [10][11]. Furthermore, there are a number of unresolved basic questions. For example, if memory information is stored in the synapse, how is it accessed (i.e., during a memory recall)? [12]. In neuroscience literature, there is a growing number of findings against the synaptic hypothesis. For example: *"Long term memory storage and synaptic change can be dissociated"* [13]; *"Increased synaptic strength that is the result of cellular consolidation is thus not a critical requisite for storing a memory"* [14]; *"When enhanced synaptic strength between engram cells is abolished, the memory is not"* [15]; *"Memory does not reside in altered synaptic conductances."* [16]. As summarized succinctly in [17]: (we do not know) *"the physical medium in nervous tissue that is modified in order to preserve these empirical quantities for use in later computations."* These findings beg the question: if not in the synapse, then, where?

A recent groundbreaking work by Hesslow [18] has demonstrated that memory information may be stored inside the neuron (i.e., the cell-intrinsic engram) [19]. Gallistel proposed the cell-intrinsic memory hypothesis [16], suggesting that memory information resides either in the cell membrane or within the cell. The most recent conclusion is that memory is likely to be found in *"information-bearing molecules inside neurons"* [20]. In a related work, Akhlaghpour recently proposed an RNA-based theory of natural universal computation [21].

Here, we start with the hypothesis that memory information is stored inside the neuron. From there, we examine how the information might be transmitted reliably between neurons by drawing on information theory and communications systems engineering perspectives. We show that channel errors can be mitigated using a neurobiologically plausible error control coding (ECC) scheme. We further illustrate that source errors can be overcome using source redundancy configurations. We perform a simulation to demonstrate how reliable communications over noisy channels can be attained using imperfect neurons. We propose 4 alternative hypotheses for the synapse's function from a communications systems perspective. We end by highlighting some key open research questions and outlining some potential areas of application.

## II. Reliable Communication of Memory Information Between Neurons

As a starting point, we apply Shannon's classical model of a communications system [22] (see Figure 1) to neuronal communications. Here, an origination neuron serves as the information source since we are working on the basis that information is stored inside the neuron. From there, the information is modulated and transmitted over a noisy channel. In our work here, we assume Inter-spike Interval (ISI) as the neuronal modulation scheme. This scheme is known in communications systems engineering as Differential Pulse Position Modulation (DPPM) [23][24][25]. The noisy communications channel is the axon, consisting of myelin and Node of Ranvier segments that repeatedly boost/refresh the transmitted signal. At the end of the axon, the synapse demodulates the transmitted signal from its electrical form (i.e., neural spikes) into chemical form (i.e., neurotransmitters), upon which the transmitted information enters the destination neuron. We refer the reader to [20] for a molecular biology description of the likely information storage medium/mechanism.

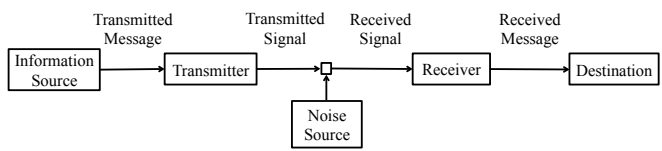

Figure 1: Classical model of a communications system [22].

Under this Shannon model, there are 2 possible sources of errors along the neuronal communications chain, to which we term here as source errors and channel errors. Source errors refer to errors that (might) arise from writing, storing and/or retrieving information in the origination neuron. Channel errors refer to errors that (might) arise from the transmission of the information from the origination neuron to the destination neuron via the noisy axonal channel. These 2 types of errors are related to Shannon's well-known theorems, namely, the source coding theorem and the channel coding theorem [26]. It is worth noting here that most neuroscientists tend to employ the source coding theorem in their analyses and modeling (i.e., maximum information compression) [27][28], whereas the channel coding theorem is almost always ignored.

*A. Resilience against channel errors*

In communications systems, there are 2 typical options for mitigating channel errors that arise from the information transmission process. One option is the employ a very high Signal-to-Noise Ratio (SNR). Another option is to employ an error control coding (ECC) scheme. The use of a very high SNR comes with a high energy/power requirement (i.e., energetically expensive). Given that the brain only consumes an estimated 12 watts of power [29], this option is unlikely to be employed. The use of an ECC scheme is more likely. An ECC scheme introduces redundancy by using more bits than minimally required to represent the information. In many ways, this is the opposite goal to source coding where maximum compression is attained using a minimum number of bits.

Almost all types of ECC schemes require buffering in the encoding and/or decoding process. For example, in a (7,4,3) Hamming code, a buffer of 4 (information) bits is required at the encoder in order to generate the 3 redundancy (i.e., parity) bits, while a 7-bit buffer is required at the decoder in order to perform error correction [30]. Many high performance ECC schemes require a relatively long block size (e.g., 65,536 bits) [31][32]. One major problem of large block sizes is that the large buffers incur time delays. Meaning, all the bits of a codeword (or the entire code block) must arrive before the encoding and/or decoding process can begin. This buffer delay slows down the information processing and limits the maximum processing speed, which can potentially be problematic for low latency processes in the brain. For example, the visual system processes light, which may not necessarily require light-speed processing per se, but nevertheless, require very fast processing without being slowed down (too much) by buffers. Moreover, a buffer management mechanism would be required in order to control the length of time that the buffer is held for, and to flush/empty/replace the buffer (e.g., after the encoding and/or decoding processes are completed). For these reasons, we will employ zero buffering in our work here, which offers the minimum possible latency, and therefore, the maximum possible speed of communications.

The only ECC scheme that does not require buffering at both the encoder and decoder is a rate 1/n parallel repetition code where the information bit is replicated n times, using n parallel axons that run from the originating neuron to the destination neuron. Figure 2 depicts a rate 1/5 parallel repetition code, in contrast to the no ECC scenario. In such a rate 1/5 ECC scheme, there are 4 redundant bits for every information bit. A rate 1/n parallel repetition code is also the easiest possible ECC scheme to decode [26]. For soft-decision decoding, the only mathematical operation required is to take the average of all n modulation signals [26]. Reliable transmission can be achieved with a low transmission error at a relatively low SNR. It also offers the lowest possible latency, which means the fastest possible communications and information processing. The parallel repetition code is neurobiologically plausible given the abundance of parallel nerve fibers in the brain.

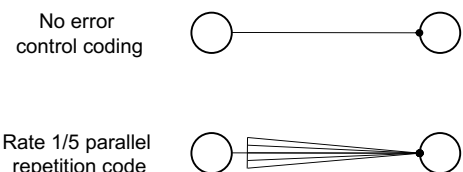

Figure 2: No ECC, and a rate 1/5 parallel repetition code.

*B. Resilience against source errors*

Recall that the source is the neuron itself where memory information is stored. The single-source point-to-point configuration previously discussed in Figure 2 is also known as a Single-Input and Single-Output (SISO) configuration [33], where a single neuron serves as input into the communications channel, with a single destination neuron receiving the output (see Figure 3). A single source configuration lacks resilience to errors because, if the source is erroneous (e.g., corrupted), then, the ECC scheme would no longer be helpful since the error would be replicated n times using the parallel repetition code axons and erroneous information will be decoded at the destination neuron. That is, a single source configuration presents the problem of a single point of failure. This vulnerability and susceptibility to error is unlikely to be acceptable in the brain. It would, therefore, make sense to have redundant sources (i.e., extra copies of the same source information), akin to a RAID system (Redundant Array of Independent Disks) in computer storage [34]. We outline here 3 possible source redundancy configurations.

*1) Multiple-Input and Single-Output (MISO)*

The most obvious solution to the problem of source error is to have multiple identical sources. Thus, instead of the information being stored in just one source neuron, the same information would be stored in multiple source neurons. This way, if one source neuron is erroneous, the remaining source neurons would still possess and transmit the correct information to the destination neuron. Such a redundant configuration is known as Multiple-Input and Single-Output (MISO) [33], as depicted in Figure 3. It is worth noting here that each of the source neurons can have either only 1 axon connecting it to the destination neuron (i.e., no ECC scheme) or n parallel axons (connecting it to the destination neuron) resulting in mix of ECC and source redundancies. Regardless of 1 or n axons per source neuron, the decoding process at the destination neuron remains the same (i.e., compute the average of all input signals).

*2) Single-Input and Multiple-Output (SIMO)*

Similar to the case of a potential error using a single source neuron, a potential error can also happen at a single destination neuron. For example, an error may arise at the sole destination neuron during the decoding process. If that happens, then, the error could further propagate from this destination neuron (which now becomes the source neuron) to other destination neurons in a catastrophic chain-effect manner. A solution to this problem is to transmit the information from one source neuron to multiple (identical) destination neurons, so that there is no single point of (destination) failure. Such a configuration is known as Single-Input and Multiple-Output (SIMO) [33], as depicted in Figure 3. Conceptually, SIMO is similar to a broadcast system.

*3) Multiple-Input and Multiple-Output (MIMO)*

A MISO configuration can be combined with a SIMO configuration in order to attain both source and destination redundancies. Such a scheme is known as a Multiple-Input and Multiple-Output (MIMO) configuration [33], as depicted in Figure 3. The MIMO configuration is a highly power efficient system [35] that is currently used in the most advanced communications systems such as 5G [36] and 802.11ax Wi-Fi networks [37]. This power efficiency could plausibly reflect the brain's low power consumption [29]. MIMO configurations have also been used to develop a cognitive prosthesis [38].

## III. SIMULATION OF RELIABLE COMMUNICATIONS OVER NOISY CHANNELS USING IMPERFECT NEURONS

In this section, we show how reliable communications can be achieved in noisy channels using imperfect neurons. We employ the electrophysiology experimental data from [18] and directly utilize findings that were reported in [24] as the starting point. From there, we produce a MISO configuration simulation that mimics the experimental data to show that reliable communication is possible.

### A. Experimental setup and key findings

Individual cerebellar Purkinje cell neurons in ferrets have recently been found to respond to an artificial conditioned stimulus (CS) [16][18]. Prior to any training, each neuron fired at its usual patterns and reacted indifferently to the artificial stimulus. After training, each neuron was able to recognize the artificial stimulus by pausing its firing for a time duration that is approximately the same as the active duration of the artificial stimulus, as shown in the raster plot in Figure 4. Each of the black dots represents a neural spike. The 2 vertical blue lines signify the beginning and the end of the 200 ms artificial stimulus. The x-axis is the time stamp, with time = 0 signifying the beginning of the artificial stimulus. The y-axis represents the 20 post-training trials, and the horizontal lines of black dots are the neural firing patterns for each of the 20 corresponding trials. The key finding [18][24] is that the neuron pauses its firing for approximately (but not exactly) the time duration (i.e., ~200 ms) that the artificial stimulus is active. The accuracy of the spike pauses varies from one trial to another. Figure 5 shows the histogram of the 20 spike pauses, with a mean of 192.5 ms and a mode of 200 ms. As reported in [24], it is plausible that the neuron intended to pause for the same duration as the artificial stimulus (i.e., 200 ms), but was unable to do so accurately and exactly every time due to its neurobiological imperfections, along with noisy channels.

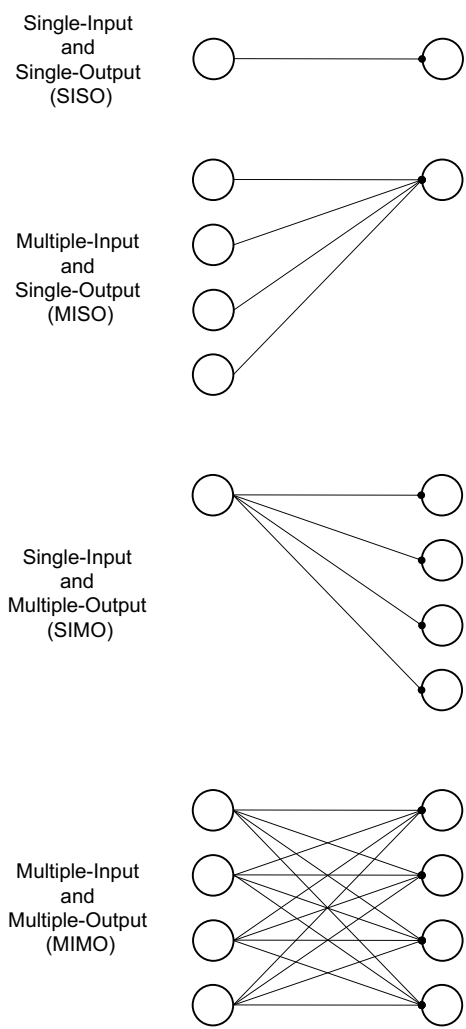

Figure 3: SISO, MISO, SIMO and MIMO configurations.

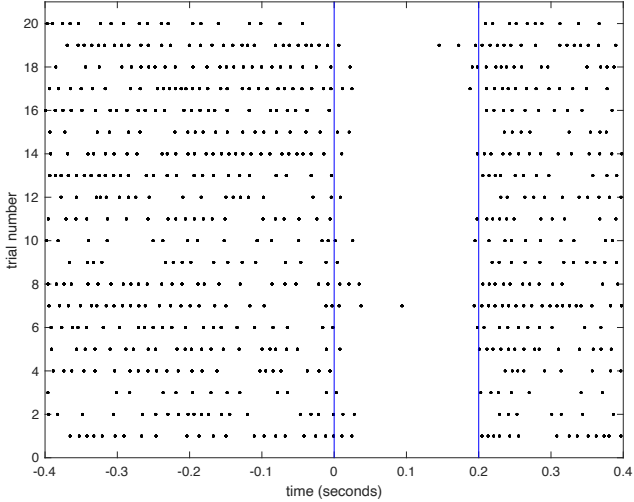

Figure 4: Raster plot of a neuron that was trained to learn a 200 ms artificial stimulus [24].

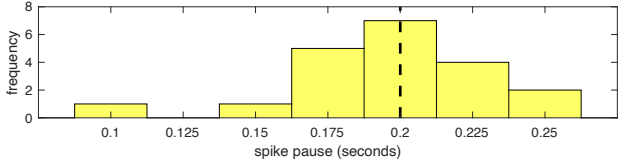

Figure 5: Histogram of the 20 spike pauses [24].

### B. A MISO source redundancy configuration

We performed a simulation by applying a MISO configuration whereby $n$ neurons were employed in parallel to perform the same task (of recognizing the 200 ms artificial stimulus), as shown in Figure 6. For convenience, we assumed that the spike pauses have a Gaussian distribution, and that each of the $n$ neurons has the same mean (i.e., 200 ms) and variance (i.e., $\sigma_1 = \sigma_2 = \ldots = \sigma_n$). We note that our approach here can be extended to other statistical models of noise. We also note here that this MISO configuration can also be used to decode a parallel repetition code at the destination neuron.

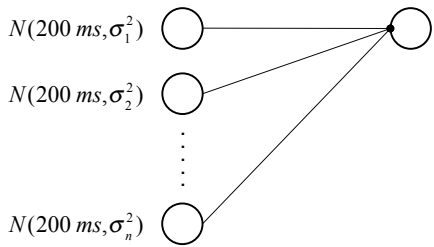

Figure 6: A MISO configuration applied to the data from [24].

Figure 7 shows the simulation results for cases of $n = 1$, 5 and 50. We chose $\sigma^2 = 0.0012$ such that the distribution of each neuron's spike pause is similar to that found in the experimental data. For $n = 1$ (i.e., top graph), the histogram resembles the one in Figure 5, as expected. For $n = 5$ (i.e., middle graph), the variance of the spike pause distribution reduces, thereby increasing the reliability of information transmission. For $n = 50$, the reliability is shown to improve further. What we have demonstrated here is that the use of multiple lower reliability neurons in a MISO configuration to perform the same task will significantly increase the overall reliability when combined at the destination neuron.

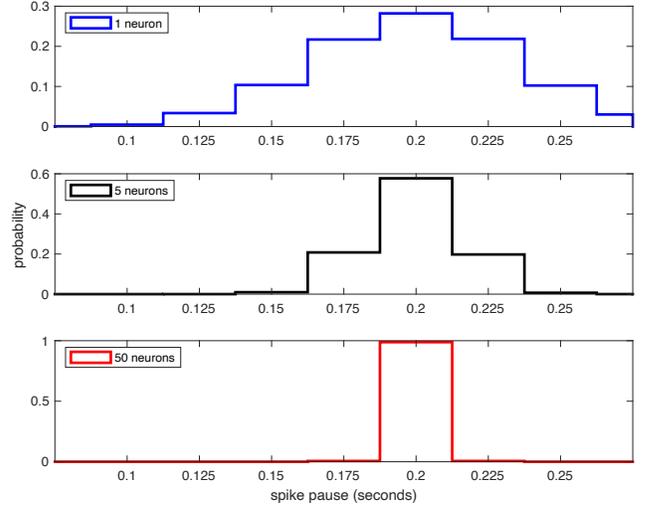

Figure 7: Effect of increasing the number of neurons in a MISO configuration to increase overall reliability.

Figure 8 shows the reduction in normalized variance with increasing number of neurons in a MISO configuration. Analytically, the overall variance can be calculated using a harmonic mean formula that is similar to that used for calculating parallel resistance in an electrical circuit:

$$\frac{1}{\sigma_{MISO}} = \frac{1}{\sigma_1} + \frac{1}{\sigma_2} + \ldots + \frac{1}{\sigma_n}$$

In our simulation, the variance of each parallel neuron is identical, and therefore, the resulting overall variance is:

$$\sigma_{MISO} = \frac{\sigma_1}{n}$$

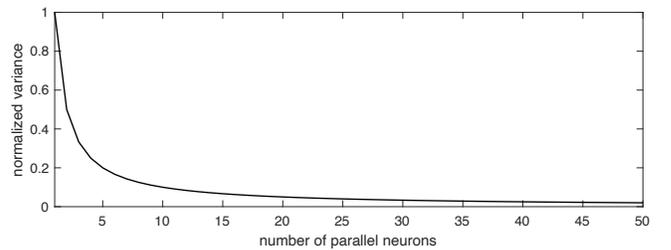

Figure 8: Reduction of variance with increasing number of neurons in a MISO configuration.

## IV. PLAUSIBLE ALTERNATIVE HYPOTHESES FOR THE SYNAPSE'S FUNCTION

In this section, we present a few plausible alternative hypotheses for the synapse's function from the perspectives of electrical circuits and communications systems engineering. These hypotheses are compatible with the existing notion of a synaptic weight in neuroscience, with the exception that the weight is used for communications instead of computations.

### A. Synapses preventing a voltage overload

From a pure electrical circuit standpoint, it is plausible that the neurons' synapses are necessary in order to avoid a voltage overload. For example, suppose that 50 source neurons are connected to one destination neuron in a MISO configuration, with each neural spike signal having a peak voltage of 40 mV. By linear superposition, the summation of these 50 neural spikes could result in a maximum peak voltage of 50 x 40 mV = 2000 mV, which would quite likely overload and damage the destination neuron. Given that each neuron in the brain is estimated to have about 7,000 synaptic connections with other neurons [39], this potential overload problem is not trivial. The synapse (at the destination neuron) converts electrical neural spike signals into chemical reactions (e.g., by releasing neurotransmitters), thereby, preventing such a voltage overload. From this electrical point of view, the synapse is a crucial, requisite and indispensable part of the brain.

### B. Synapses performing part of the averaging operation

Averaging (as a mathematical operation) is an instrumental part of the decoder for a rate 1/n parallel repetition code. A neuron's synaptic weight could plausibly perform part of this averaging operation by weighting (or multiplying) each input signal by a 1/n scalar, such that the destination neuron only needs to sum all these weighted signals in order to complete the averaging operation. In such a situation, each weight would be equal/identical, as demonstrated in the simulation presented in the previous section.

### C. Synapses as reliability weights for redundant inputs

In a source redundancy configuration such as MISO (Figure 3), it is plausible that some inputs may be more reliable than others. This difference could arise because some source neurons may be more reliable than others. In addition, some axonal channel pathways may be noisier than others. Consequently, at the destination neuron, it would make sense for a higher weighting to be applied to a more reliable input prior to being combined. The synapse could plausible be performing such a proportionate weighting of reliability.

### D. Synapses compensating for channel effects

In wireless radio communications, multiple copies of the same transmitted radio signal travel through a different pathway en route to the destination/receiver, as illustrated in Figure 9. One copy of the radio signal might arrive directly at the destination/receiver; another copy of the radio signal might bounce off a house before arriving at the destination/receiver; another copy of the radio signal might bounce off a moving car. At the destination/receiver, each copy of the radio signal would have a slightly different characteristic – specifically, different signal strength (i.e., SNR) and different arrival times (due to a different length of physical path travelled). For example, the signal that bounced off the house is weaker than the direct path signal, and it has also travelled a longer distance. This problem is known as multipath propagation [30]. A solution to this multipath propagation problem is the RAKE receiver [30][40] where each of the multipath signals is compensated separately for their different signal strength and arrival delays prior to being combined together.

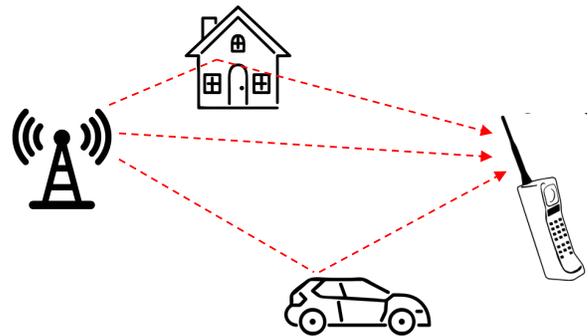

Figure 9: Multipath propagation in wireless radio communications.

In the brain, a situation similar to multipath propagation can plausibly exists. For example, if the destination neuron were to receive inputs from M source neurons (or M parallel repetition pathways), each of the M axons may be similar but not identical in terms of communications channel characteristics (e.g., slightly different lengths, slightly different number of myelin and Node of Ranvier segments, slightly different conductive characteristics). Consequently, the M inputs could each have a slightly different neural spike voltage and arrival time. The synapse may therefore plausibly function as a weighted compensator, individually tuned to account for each signal strength and arrival delay variations, so that all the neural spikes arrive at the destination neuron within an acceptable synchronized time window/margin.

In a wireless radio receiver, each of the channel estimators in the RAKE receiver has to be updated quite frequently, in order to account for and adapt to the fast-changing communications channel characteristics. On the other hand, in the brain, the channel characteristics (i.e., the M axons) are relatively constant (i.e., unchanged), which means that, once a synapse has been adapted or calibrated for that particular axon's channel characteristics, it may not have to be changed again, at least not as significantly or as frequently.

## V. Conclusions and Discussions

Starting with the assumption that memory information is stored inside the neuron instead of in the synapse, we applied Shannon's model to examine how information might be transmitted reliably between neurons. We showed that channel errors can be mitigated by using neurobiological plausible parallel repetition codes. We further illustrated that source errors can be overcome using MISO, SIMO and MIMO redundancy configurations. Based on a set of groundbreaking electrophysiology data [18], our simulation demonstrated that reliable communications over noisy channels using imperfect neurons are feasible. From a communications systems engineering perspective, we put forward 4 plausible alternative hypotheses for the synapse's function that are compatible with existing notions of a synaptic weight.

It is obvious that we don't have all the answers here. In fact, far from it. Our intention is to outline the foundational groundwork to catalyze a fresh look at understanding and modelling of the brain by the information theory and communications systems engineering research community. Here are some key open research questions:

- What is the size of the signal constellation of the DPPM (i.e., Inter-spike Interval) scheme? Since DNA operates on a base-4 numbering system [41], it would make logical sense for the signal constellation to be compatible, possessing at least 4 distinct modulation symbols (e.g., 4-DPPM, 16-DPPM).
- How do we better analyze the effects of noise along the axonal channel on the performance of the DPPM scheme? Specifically, what are the symbol error rate (or bit error rate) versus SNR performance tradeoffs?
- If two successive modulation symbols (i.e., neural spikes) are too close together, both symbols could overlap and produce a signal disruption known as Intersymbol Interference [30], whereby one symbol can potentially attenuate (or cancel out) the other symbol. Instead of analyzing post-thresholded neural spike data, it could be more insightful to analyze the raw neural waveforms (i.e., the DPPM baseband modulation signals) to understand its spectral properties and identify potential effects of Intersymbol Interference.
- The NIH Brain Initiative has previously estimated that there are at least 754 different types of neurons in the human brain [42]. A prominent neuroscientist, Christof Koch, further estimated that there are *"probably thousands"* of different types [43]. How do these different types of neurons vary in terms of communications properties and characteristics (e.g., baseband pulse shape)?
- Synchronization of modulation symbols [30] is crucial when combining signals from multiple transmission channels (e.g., multipath propagation, or multiple parallel axons). How is synchronization attained and sustained in the brain [44]?

In terms of potential areas of application, one relevant area is in research on Alzheimer's disease, which is known to disrupt communication among neurons, causing impairment in function and cell death [39]. A better understanding of dissociative amnesia arising from trauma [45] could also be attained by taking into account communications systems engineering perspectives. Neurorehabilitation [46] is another area that could benefit from the concepts outlined in this paper.

## Acknowledgment

J.T. thanks F. Johansson of the Hesslow Lab at the Department of Experimental Medical Science at Lund University (Sweden) for sharing his experimental data.

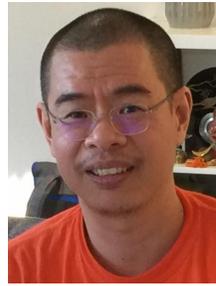

James Tee (M'17) completed his Ph.D. in Electrical & Electronic Engineering at the University of Canterbury in 2001, where he worked on Turbo Codes under the supervision of Des Taylor. Subsequently, he held various industry and policy positions at Vodafone Group, the World Economic Forum, New Zealand's Ministry of Agriculture & Forestry, and the United Nations. To facilitate his career transitions, he pursued numerous supplementary trainings, including an MBA at the Henley Business School, and an MPhil in Economics (Environmental) at the University of Waikato. In 2012, James began his transition into scientific research at New York University (NYU), during which he completed an MA in Psychology (Cognition & Perception) and a PhD in Experimental Psychology (Neuroeconomics). Afterwards, he worked as an Adjunct Assistant Professor at NYU's Department of Psychology, and a Research Scientist (Cognitive Neuroscience) at Quantized Mind LLC. Since 2017, he is an Adjunct Research Fellow at the University of Canterbury. James is a New York State licensed acupuncturist, with an MS in Acupuncture from Pacific College of Oriental Medicine. In September 2020, he began further training in Substance Abuse Counseling at the New School for Social Research (NSSR). His current research interests in neuroscience focuses on the communications and information theoretic aspects of the brain. He is also currently pursuing research on the integration of acupuncture and psychotherapy for treating mental health disorders.

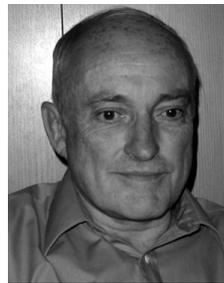

Desmond P. Taylor (LF'06) received the Ph.D. degree in electrical engineering from McMaster University, Hamilton, ON, Canada, in 1972. From 1972 to 1992, he was with the Communications Research Laboratory and the Department of Electrical Engineering, McMaster University. In 1992, he joined the University of Canterbury, Christchurch, New Zealand, as the Tait Professor of communications. He has authored approximately 250 published papers and holds several patents in spread spectrum and ultra-wideband radio systems. His research is centered on digital wireless communications systems focused on robust, bandwidth-efficient modulation and coding techniques, and the development of iterative algorithms for joint equalization and decoding on fading, and dispersive channels. Secondary interests include problems in synchronization, multiple access, and networking. He is a Fellow of the Royal Society of New Zealand, the Engineering Institute of Canada, and the Institute of Professional Engineers of New Zealand.